\documentclass[a4paper,twoside,10pt]{letter}
\usepackage{saj,graphicx,multicol,subeqnarray}

\def\papertype{Original Scientific Paper}

\newcommand{\SII}{[S\,{\sc ii}]}
\newcommand{\OIII}{[O\,{\sc iii}]}
\newcommand{\Halpha}{H${\alpha}$}
\newcommand{\D}{$^\circ$}
\def\p0{\phantom{0}}
\def\it{\sl}

\def\degr{\hbox{$^\circ$}}
\def\arcmin{\hbox{$^\prime$}}
\def\arcsec{\hbox{$^{\prime\prime}$}}
\def\SNR{\mbox{{SNR~J0527--6549}}}
\def\udc{524.354--77 : 524.722.3}
\begin{document}
\baselineskip=3.1truemm
\columnsep=.5truecm
\newenvironment{lefteqnarray}{\arraycolsep=0pt\begin{eqnarray}}
{\end{eqnarray}\protect\aftergroup\ignorespaces}
\newenvironment{lefteqnarray*}{\arraycolsep=0pt\begin{eqnarray*}}
{\end{eqnarray*}\protect\aftergroup\ignorespaces}
\newenvironment{leftsubeqnarray}{\arraycolsep=0pt\begin{subeqnarray}}
{\end{subeqnarray}\protect\aftergroup\ignorespaces}
%


\markboth{\eightrm MULTIFREQUENCY RADIO OBSERVATIONS OF \SNR}
{\eightrm L.M. Bozzetto, M.D. Filipovi\'c, E.J. Crawford, A. Mendik, B. Wardlaw, J.L. Payne, A.Y. De Horta and I.S. Boji{\v c}i{\'c} }

{\ }

\publ

\type

{\ }


\title{MULTIFREQUENCY RADIO OBSERVATIONS OF A SNR IN THE LMC. \\ The case of \SNR\ (DEM L204)} 


\authors{L.M. Bozzetto$^1$, M.D.~Filipovi\'c$^1$, E.J. Crawford$^1$, I.S. Boji{\v c}i{\'c}$^1$, J.L. Payne$^1$}\authors{A. Mendik$^1$, B. Wardlaw$^1$ and A.Y. De Horta$^1$}

\vskip3mm


\address{$^1$School of Computing and Mathematics, University of Western 
Sydney
\break Locked Bag 1797, Penrith South DC, NSW 1797, Australia}  

\Email{m.filipovic@uws.edu.au}


\dates{March 2010}{March 2010}



\summary{We present a detailed study and results of new Australia Telescope Compact Array (ATCA) observations of supernova remnant, \SNR. This Large Magellanic Cloud (LMC) object follows a typical supernova remnant (SNR) horseshoe morphology with a diameter of D=(66$\times$58)$\pm$1~pc which is among the largest SNRs in the LMC. Its relatively large size indicates older age while a steeper than expected radio spectral index of $\alpha$=--0.92$\pm$0.11 is more typical for younger and energetic SNRs. Also, we report detections of regions with a high order of polarization at a peak value of $\sim$54\%$\pm$17\% at 6~cm.}



\begin{multicols}{2}
{


\section{1. INTRODUCTION}


The Large Magellanic Cloud (LMC) contains one of the most vigorous star forming regions, (such as 30~Dor) in our Local Group of galaxies. Located at a distance of 50~kpc (Di Benedetto 2008), it is one of the best galaxies to study supernova remnants (SNRs), due to the favourable position in the direction toward the South Pole. As well as its viewing position, the LMC is also located in one of the coldest areas of the radio sky, which allows us to observe radio emission without the interruption from Galactic foreground radiation. In addition to this, the LMC resides outside of the Galactic plane and therefore the influence of dust, gas and stars is negligible. 

Predominately non-thermal emission is a well-known characteristic of SNRs in the radio-continuum. Although SNRs have a typical radio spectral index of $\alpha\sim-0.5$ defined by $S\propto\nu^\alpha$, this can significantly change, due to the fact of the wide variety of types of SNRs in various environments (Filipovi\'c et al.~1998a). The ISM's morphology, structure, behaviour and evolution can be attributed to SNRs, and in turn this heavily impacts the evolution of SNRs, as they are dependant on the environment in which they reside. 

Here, we report on new radio-continuum and optical observations of previously poorly studied \SNR. The observations, data reduction and imaging techniques are described in Section~2. The astrophysical interpretation of newly obtained moderate-resolution total intensity and polarimetric image in combination with the existing Magellanic Cloud Emission Line Survey (MCELS) images are discussed in Section~3.

\section{2. OBSERVATIONAL DATA}

\subsection{2.1. Previous observations of \SNR}

\SNR\ was initially classified as an SNR based on the Einstein X-ray survey by Long et al.\ (1981) (named LHG~39). Mathewson et al. (1983) catalogued \SNR\ based on their optical observations, reporting an estimated optical size of $237\arcsec\times212\arcsec$ (57$\times$51$\pm$1~pc; using 50~kpc as the distance to the LMC). They also studied this object using MOlonglo Synthesis Telescope (MOST) survey. This SNR showed in Clarke et al.~(1976) 408~MHz MC4 catalogue as a distinctive point-like radio source which integrated flux density was later re-measured by Mathewson et al. (1983) to be of 260~mJy. Mills et al.~(1984) detected this source with specific MOST pointings and indicated a spectral index of $\alpha$=--0.45. However, Filipovi\'c et al. (1998b) reported a flatter spectral index of $\alpha$=--0.2$\pm$0.1. 

An optically identified object at this position was also listed in the Davies et al.~(1976) catalogue of nebular complexes the Magellanic Clouds as emission nebulae --- DEM\,L204. Chu and Kennicutt~(1988) classified \SNR\ to belong to Population~II? (with '?' indicating uncertain classification) group with very distant (non-influential) stelar association of LH\,53 at some 340~pc. Filipovi\'c et al.~(1998b), using {\it ROSAT} All Sky Survey (RASS) observations, detected X-ray emission from \SNR\ (LMC RASS~213). Filipovi\'c et al. (1998a) added further confirmation, with a set of single dish Parkes radio-continuum observations on a wide frequency range (Filipovi\'c~et~al.~1995, 1996). Blair et al.~(2006) reported marginal detection only in C {\sc iii} at far ultraviolet wavelengths based on FUSE (Far Ultraviolet Spectroscopic Explorer) satellite. Finally, Haberl and Pietsch (1999) (named SNR as HP~180) discuss the X-Ray properties of \SNR\ based on ROSAT PSPC observations. Most recently Payne et al. (2008) presented optical spectroscopy of a wide range of LMC SNRs including \SNR. They found an enhanced [S\textsc{ii}]/H$_\alpha$ ratio of 0.8 typical for SNRs. 

\subsection{2.2. New observations of \SNR}

We observed \SNR\ with the Australia Telescope Compact Array (ATCA) on 2$^\mathrm{nd}$ October 1997, using the array configuration EW375, at wavelengths of 3 and 6~cm ($\nu$=8640 and 4800~MHz). Baselines formed with the $6^\mathrm{th}$ ATCA antenna were excluded, as the other five antennas were arranged in a compact configuration. The observations were carried out in the so called ``snap-shot'' mode, totaling $\sim$1 hour of integration over a 12 hour period. Source PKS~B1934-638 was used for primary calibration and source PKS~B0530-727 was used for secondary (phase) calibration. The \textsc{miriad} (Sault and Killeen~2010) and \textsc{karma} (Gooch~2006) software packages were used for reduction and analysis. More information on the observing procedure and other sources observed in this session/project can be found in Boji\v{c}i\'c~et~al.~(2007), Crawford~et~al.~(2008a,b; 2010) and \v{C}ajko~et~al.~(2009).

Images were formed using \textsc{miriad} multi-frequency synthesis (Sault and Wieringa~1994) and natural weighting. They were deconvolved using the {\sc clean} and {\sc restor} algorithms with primary beam correction applied using the {\sc linmos} task. A similar procedure was used for both \textit{U} and \textit{Q} Stokes parameter maps. Because of the low dynamic range (signal to noise ratio between the source flux and $3\sigma$ noise level) self-calibration could not be applied. The 6~cm image (Fig.~1) has a resolution of 41.4\arcsec$\times$30.3\arcsec\ at PA=0\D\ and an estimated r.m.s. noise of 0.15 mJy/beam. Similarly, we made an image of \SNR\ at 3~cm (Fig.~1) with resolution of 22.9\arcsec$\times$16.5\arcsec\ (PA=0\D).

We also used the Magellanic Cloud Emission Line Survey (MCELS) that was carried out with the 0.6~m University of Michigan/CTIO Curtis Schmidt telescope, equipped with a SITE $2048 \times 2048$\ CCD, which gave a field of 1.35\degr\ at a scale of 2.4\arcsec\,pixel$^{-1}$. Both the LMC and SMC were mapped in narrow bands corresponding to \Halpha, \OIII\ ($\lambda$=5007\,\AA), and \SII\ ($\lambda$=6716,\,6731\,\AA), plus matched red and green continuum bands that are used primarily to subtract most of the stars from the images to reveal the full extent of the faint diffuse emission. All the data have been flux-calibrated and assembled into mosaic images, a small section of which is shown in Figs.~2 and 3. Further details regarding the MCELS are given by Smith et al. (2006) and at http://www.ctio.noao.edu/mcels. Here, for the first time, we present optical images of this object in combination with our new radio-continuum data.

\section{3. RESULTS AND DISCUSSION}

The remnant has a typical horseshoe morphology (Fig.~1) centered at RA(J2000)=5$^h$27$^m$54.9$^s$, DEC(J2000)=--65\degr49\arcmin49.2\arcsec\ with a measured diameter at 6~cm of 271\arcsec$\times$240\arcsec$\pm$4\arcsec\ (66$\times$58$\pm$1~pc). We used \textsc{karma} tool {\sc kpvslice} to estimate \SNR\ extension at (6~cm image) the 3$\sigma$ noise level (0.45~mJy) along the major (NE) (Fig.~4) and minor (NW) axis (PA=45\D). We note that our estimate of the major diameter is significantly larger ($\sim$34\arcsec) than previously measured by Mathewson et al. (1983). We attribute this to SNR NE extension detected in \OIII\ that was not seen in any other optical wavebands (see Fig.~3; right panel). However, our measurements are in better agreement with radio diameters (61$\times$56~pc) previously reported by Mills~et~al.~(1984). The MCELS images are showing particularly strong \OIII\ emission around the North, North East (NE) and North West (NW) part of the shell. Especially, in the NW direction where \SNR\ extends to 325\arcsec$\pm$4\arcsec\ (78$\pm$1~pc) diameter which is significantly more than at radio or other optical frequencies (Fig.~3; right panel). Also, we note a prominent H$\alpha$ emission towards the southern part of SNR that is probably causing shell brightening at that end. Overall, the optical and radio-continuum emissions follow each other.

In order to estimate spectral energy distribution for this object, we use our new integrated flux density measurements at various radio frequencies with 408~MHz measurement by Mathewson~et~al.~(1983), 843~MHz measurement by Mills~et~al.~(1984) as well as at 1400~MHz (from the mosaics presented by Filipovi\'c et al.~(2009) and Hughes~et~al.~(2007)). We list these flux density measurements at various frequencies in Table~1 and then plot \SNR\ spectral index ($\alpha$) in (Fig.~5). The overall radio-continuum spectral index of \SNR\ is unusually steep ($\alpha$=--0.92$\pm$0.11) given that this is most likely older (evolved) SNR, due to its rather large size of $\sim$66~pc. Usually, a steep gradient like this would suggest a much younger and energetic SNR. However, in this case, the steepness can be contributed to the fact of missing short spacings at higher radio-continuum frequencies (4800 and 8640 MHz) and therefore missing flux. Specifically at 3~cm (where the ATCA primary beam is $\sim$300\arcsec) this SNR edges would be positioned close to the primary beam boundary where the flux tend be significantly uncertain. We also note that this may indicate that a simple model does not accurately describe the data, and that a higher order model is needed. This is not unusual, given that several other Magellanic Clouds SNR's exhibit this ``curved'' spectra (Crawford et al. 2008b). Noting the breakdown of the power law fit at shorter wavelengths, we decomposed the spectral index estimate into two components, one ($\alpha_{1}$) between 73 and 20~cm, and the other ($\alpha_{2}$) between 6 and 3~cm. The first component, $\alpha_{1}$= --0.51$\pm$0.08 is a very good fit and typical for an SNR, whereas the second, \mbox{$\alpha_{2}$=--1.60$\pm$0.34,} is a poor fit, and indicates that non-thermal emission can be described by different populations of electrons with different energy indices. Although the low flux at 3~cm (and to a lesser extent at 6~cm) could cause the large deviations, an underestimate of up to $\sim$50\% would still lead to a ``curved'' spectrum.

}

\end{multicols}

\vskip5mm

\centerline{{\bf Table 1.} Integrated Flux Density of \SNR.}
\vskip2mm
\centerline{
\begin{tabular}{cccccl}
\hline
$\nu$ & $\lambda$ & R.M.S  & Beam Size  & S$_\mathrm{Total}$  & Reference \\
(MHz) & (cm)      & (mJy) & (\arcsec) & (mJy) &\\
\hline
408 & 73 & --- & 156$\times$156 & 260 & Mathewson et al. (1983)\\
843 & 36 & 1.5 & 43$\times$43 & 166 & Mills et al. (1984)\\
1400 & 20 & 1.5 & 45.0$\times$45.0 & 140 & This Work\\
4800 & 6  & 0.15 & 41.4$\times$30.2 & 38.4 &  This Work\\
8640 & 3  & 0.17 & 22.9$\times$16.5 & 15.0 &  This Work\\
\hline
\end{tabular}}
\vspace{0.5cm}

\begin{multicols}{2}

{
 
Such a curved spectrum, as it is shown in Fig.~5, can be explained using so-called diffuse shock acceleration (DSA) theory coupled with the effect of synchrotron losses within the finite emission region. If the thin region near the shock discontinuity is not resolved by the telescope beam, the  observed emission includes some flux from electrons which have been diffused away from the place of effective acceleration and lose a significant amount of energy via synchrotron emission. As these losses are more severe for higher energy electrons, we expect this to steepen the observed synchrotron spectrum. For details see Heavens and Meisenheimer (1987), Longair (2000, and references therein).

The linear polarization images for each frequency were created using \textit{Q} and \textit{U} parameters (Fig.~6). The 6~cm image reveals some strong linear polarization, greater than various other LMC SNRs (Boji{\v c}i{\'c} et al. 2007; Crawford et al. 2008a,b; \v{C}ajko et al. 2009; Crawford et al. 2010). The mean fractional polarisation at 6~cm was calculated using flux density and polarisation:
\begin{equation}
P=\frac{\sqrt{S_{Q}^{2}+S_{U}^{2}}}{S_{I}}\cdot 100\%
\end{equation}
\noindent where $S_{Q}, S_{U}$ and $S_{I}$ are integrated intensities for \textit{Q}, \textit{U} and \textit{I} Stokes parameters. Our estimated peak value is 54\%$\pm$17\% at 6~cm and no reliable detection at 3~cm. Along the shell there is very strong uniform polarisation coinciding with the total peak intensity located at north-west side of the shell. We note that \SNR\ exhibit one of the strongest polarisations observed so far in the LMC averaged at approximately 50\% (Fig.~6) as would be expected from non-thermal SNRs. This relatively high level of polarization is (theoretically) expected for an SNR with a radio spectrum of around or less than $-0.5$ (Rolfs and Wilson 2003). Possibly, this may indicate varied dynamics along the shell. Without reliable polarisation measurements at a second frequency we could not determine the Faraday rotation, and thus cannot deduce the magnetic field strength.

}

\end{multicols}

\centerline{\includegraphics[angle=-90,width=1\textwidth]{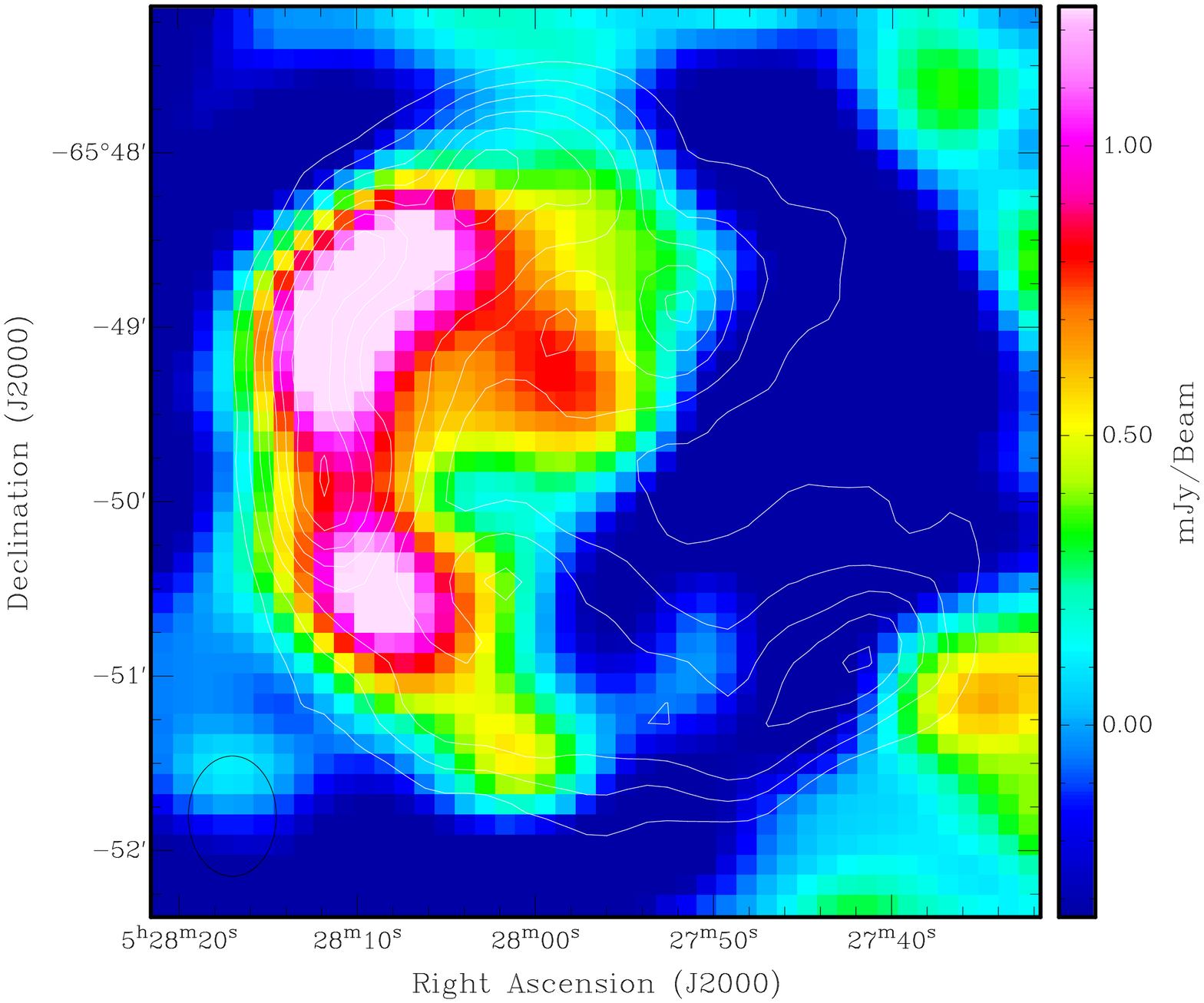}}
\figurecaption{1.}{ATCA observations of \SNR\ at 3~cm (8.6~GHz) overlaid with 6~cm (4.8~GHz) contours. The contours are 3, 5, 7, 9, 11, 13, 15 and 17$\sigma$. The black circle in the lower left corner represents the synthesised beamwidth (at 6~cm) of 41.4\,\arcsec$\times$30.3\arcsec. The sidebar quantifies the pixel map and its units are mJy/beam.
}

\centerline{\includegraphics[width=1\textwidth]{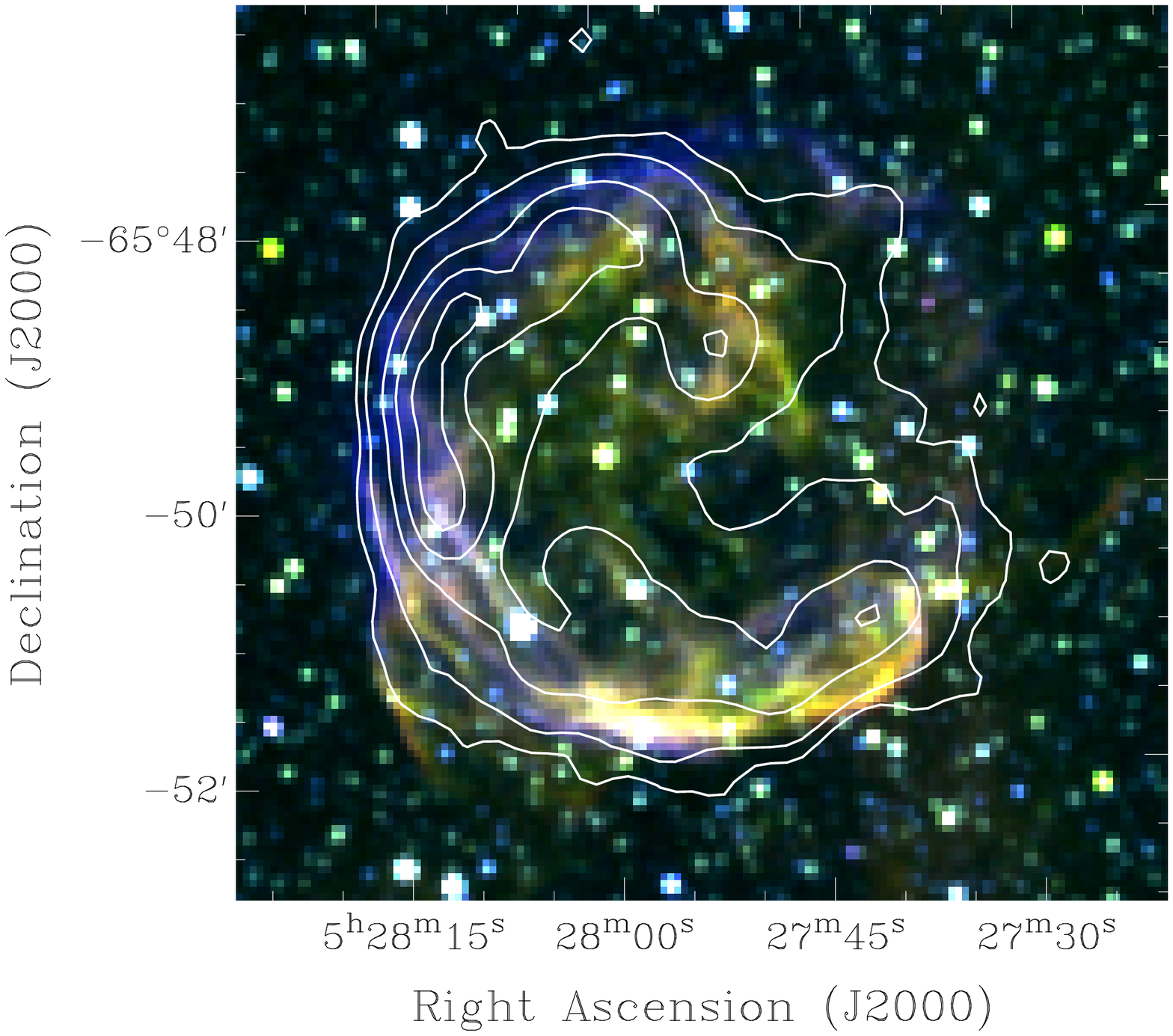}}
\figurecaption{2.}{MCELS composite optical image \textrm{(RGB =H$\alpha$,[S\textsc{ii}],[O\textsc{iii}])} of \SNR\ overlaid with 6~cm contours. The contours are 1, 3, 7, 11 and 15$\sigma$.}

\centerline{\hspace*{-1cm}\includegraphics[width=0.3\textwidth]{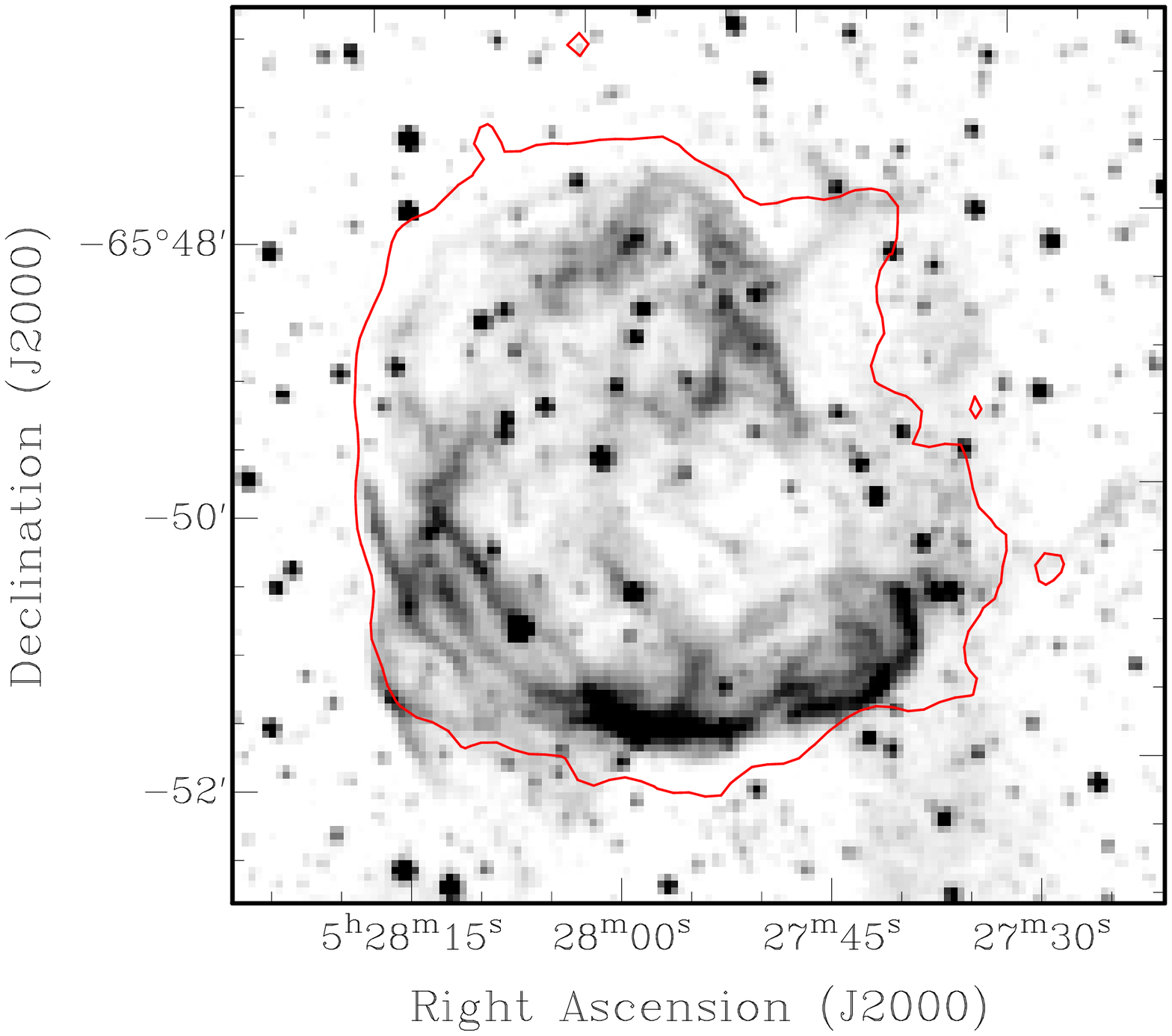}~\includegraphics[width=0.3\textwidth]{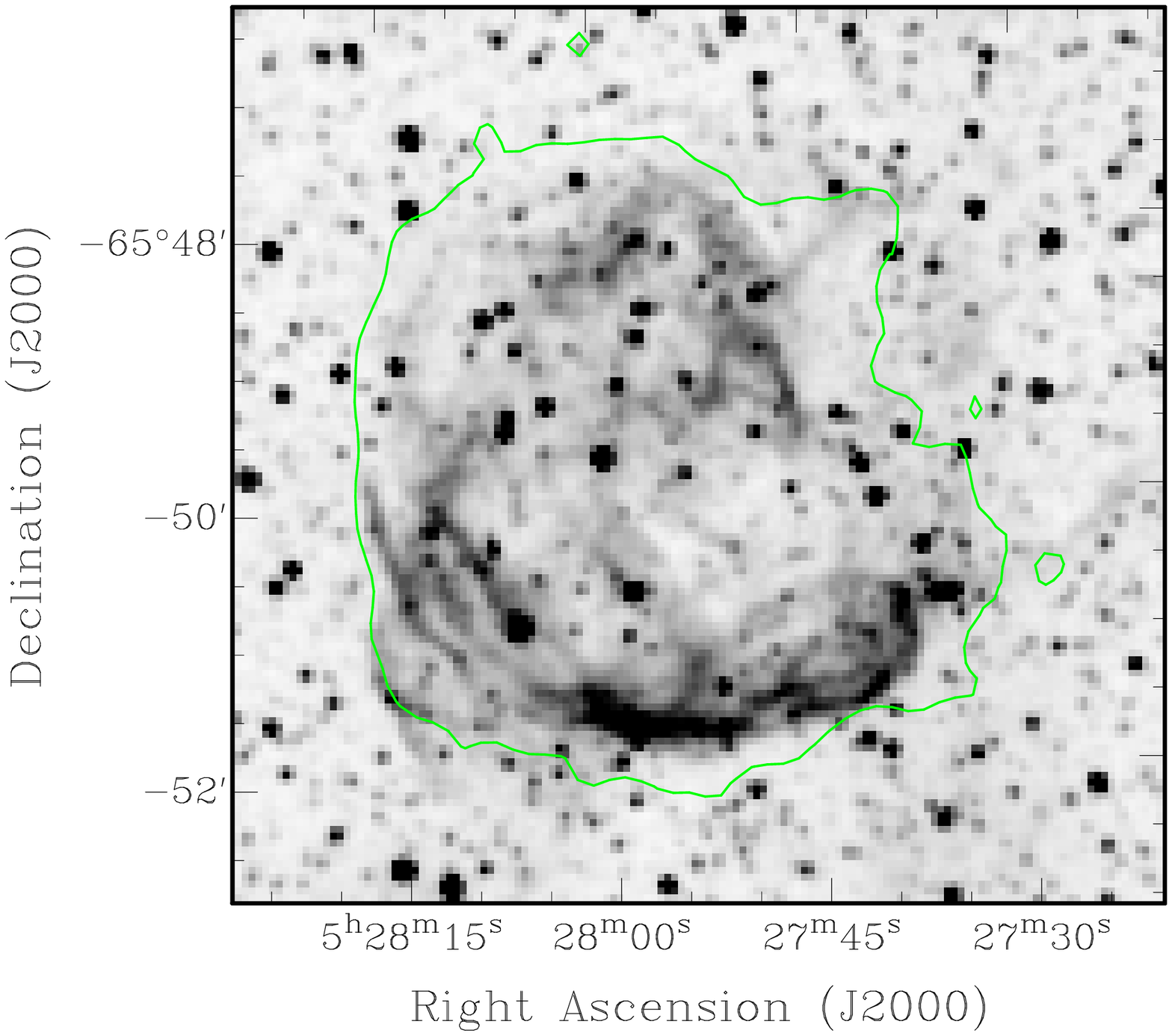}\hspace*{1cm}\includegraphics[width=0.3\textwidth]{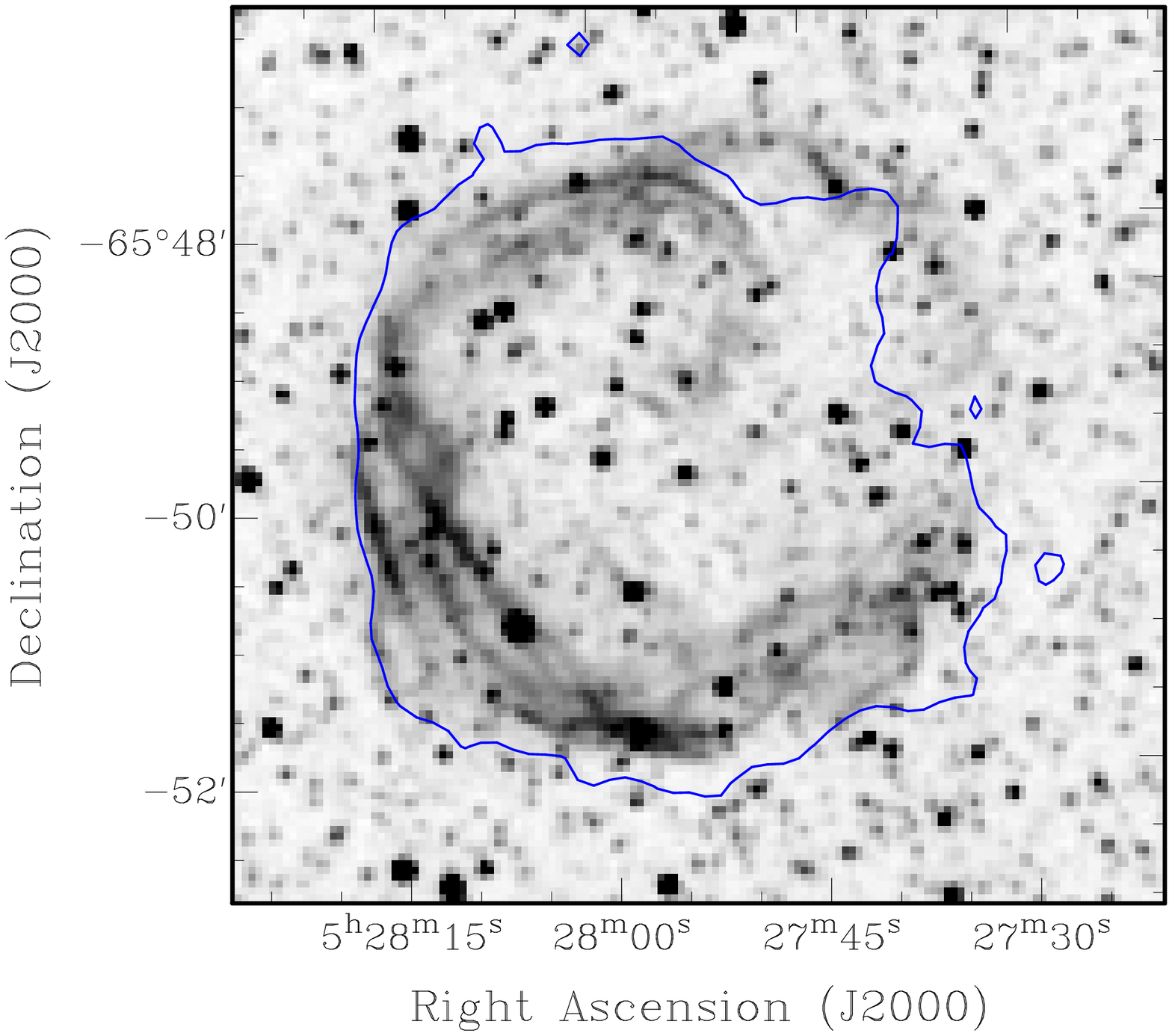}}
\figurecaption{3.}{MCELS optical images (H$\alpha$ (left), \SII\ (middle), \OIII\ (right)) of \SNR\ overlaid with a single 1$\sigma$ 6~cm contour.}

\centerline{\hspace*{-1cm}\includegraphics[width=0.3\textwidth]{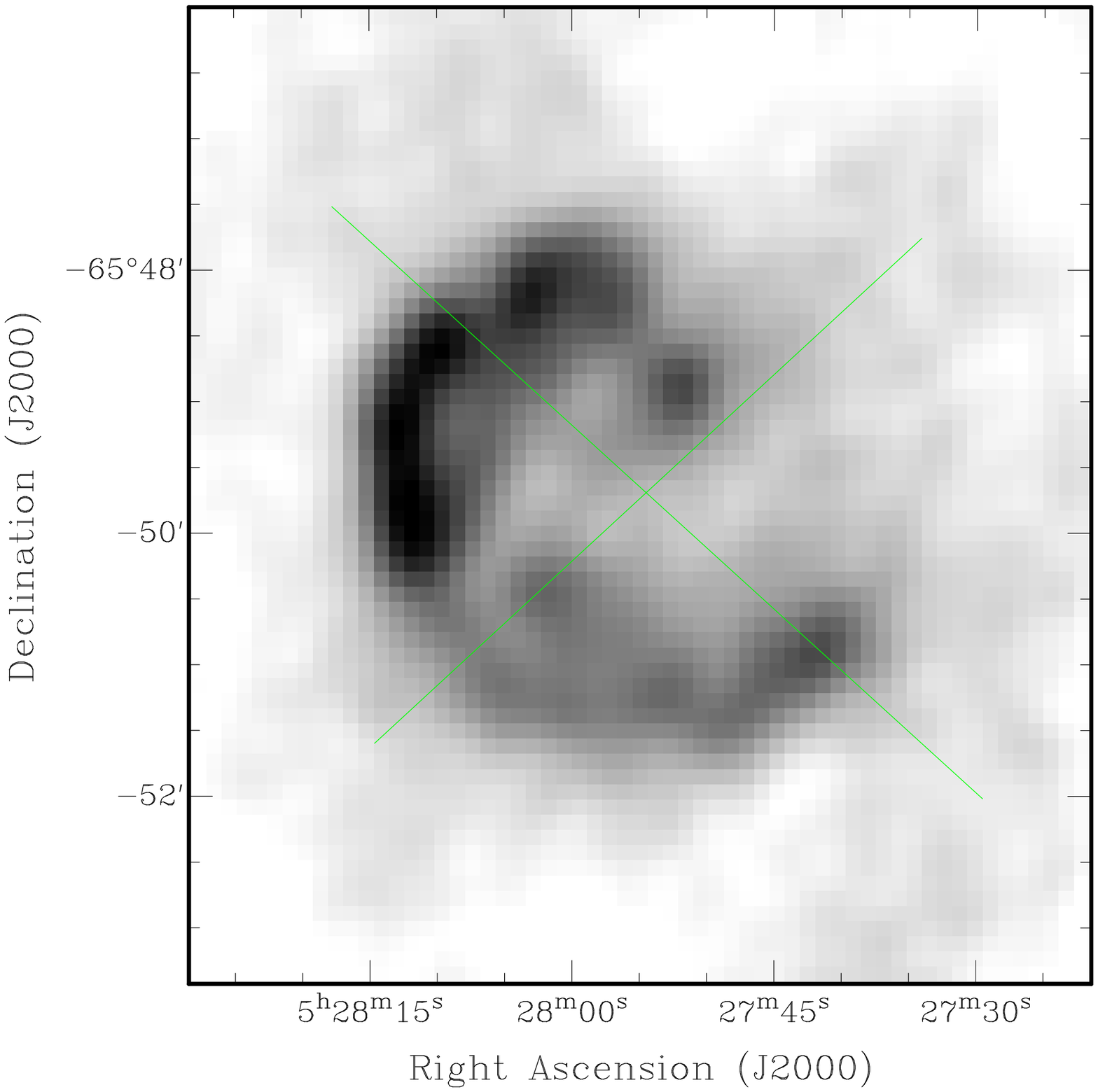}~\includegraphics[width=0.3\textwidth]{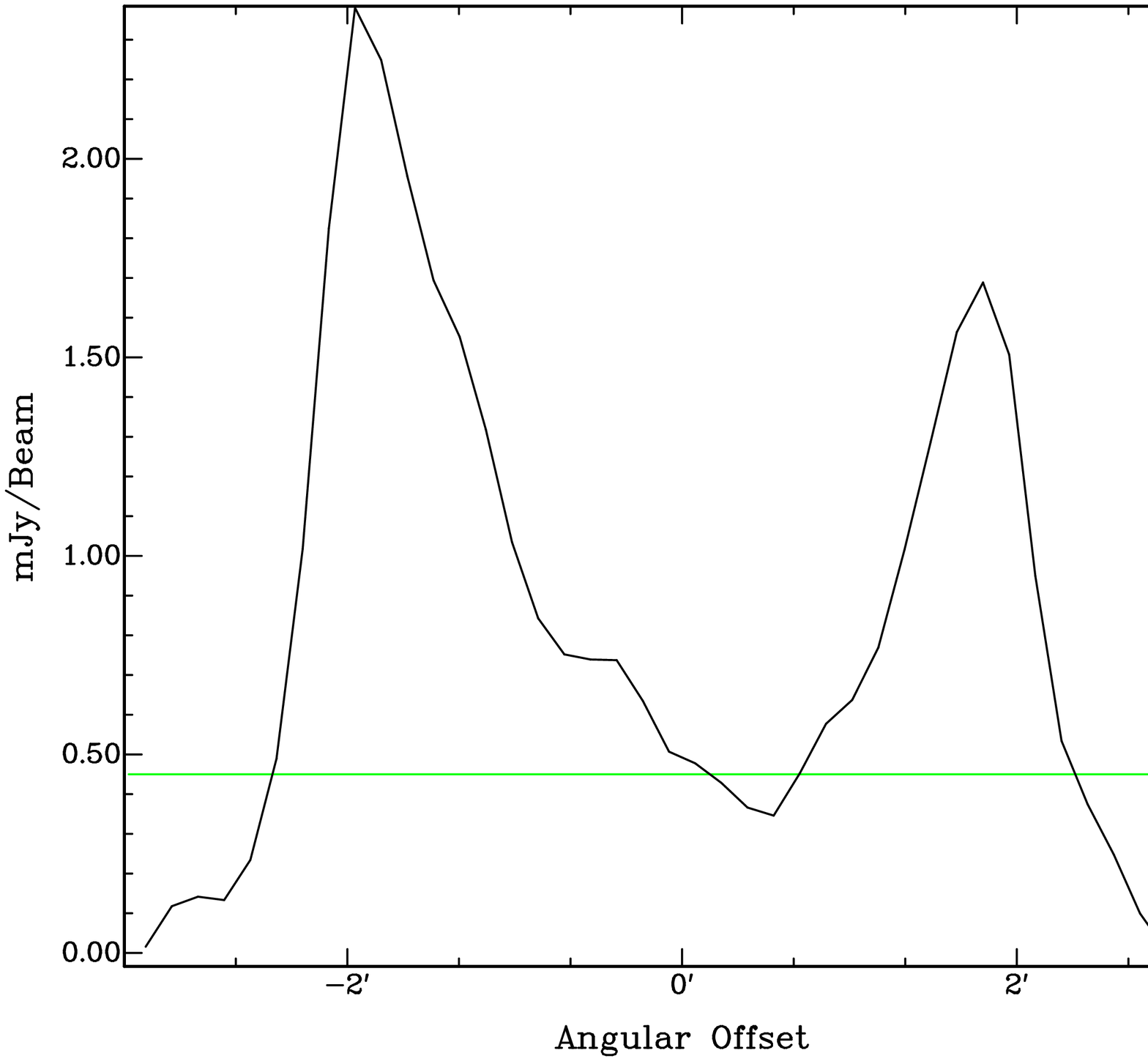}\hspace*{1cm}\includegraphics[width=0.3\textwidth]{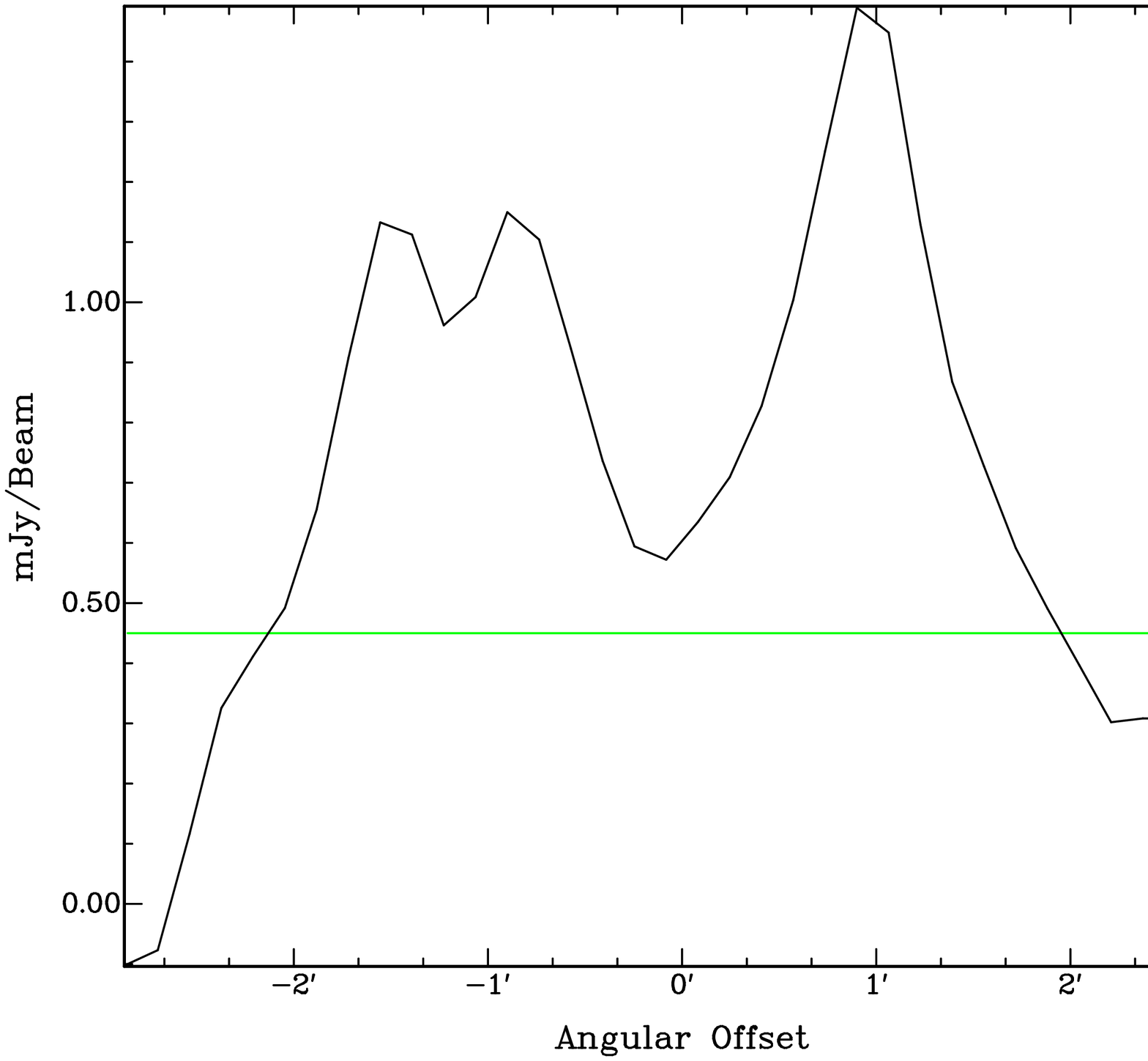}}
\figurecaption{4.}{The left image shows the major and minor axis, with the major axis starting at the NW corner and the minor axis starting at the SW corner. The center image shows the I-Profile of the major axis with the 3$\sigma$ line shown. The right image shows the I-Profile of the minor axis with the 3$\sigma$ line shown.}

\centerline{\includegraphics[angle=-90,width=.55\textwidth]{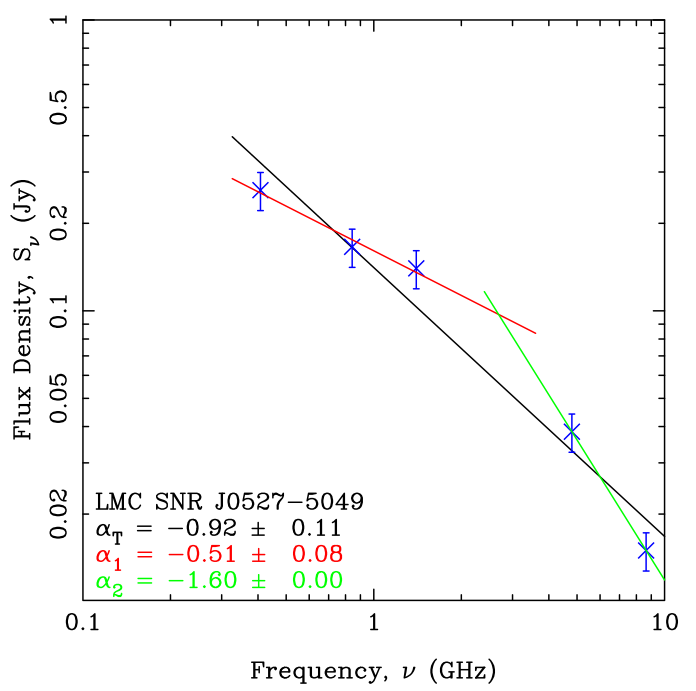} }
\figurecaption{5.}{Radio-continuum spectrum of \SNR. }

\centerline{\includegraphics[angle=-90,width=0.55\textwidth]{6cm-pol}}
\figurecaption{6.}{ATCA observations of \SNR\ at 6~cm (4.8~GHz). The blue circle in the lower left corner represents the synthesised beamwidth of 41.4\,\arcsec$\times$30.3\arcsec, and the blue line below the circle is a polarization vector of 100\%. The sidebar quantifies the pixel map and its units are Jy/beam.}

\begin{multicols}{2}

{

\section{4. CONCLUSION}

\vskip-1mm

We conducted the highest resolution radio-continuum and optical observations to date of \SNR. From this analysis, we confirmed this object as a bona-fide SNR with a relatively large diameter of 271\arcsec$\times$240\arcsec$\pm$4\arcsec\ (66$\times$58$\pm$1~pc), a complex spectral index ($\alpha$ = --0.92$\pm$0.11), strong polarisation of $\sim$54\%$\pm$17\% as well as strong, circular optical \OIII\ emission.


\acknowledgements{We used the {\sc karma} software package developed by the ATNF. The Australia Telescope Compact Array is part of the Australia Telescope which is funded by the Commonwealth of Australia for operation as a National Facility managed by CSIRO. We thank the Magellanic Clouds Emission Line Survey (MCELS) team for access to the optical images. We were granted observation time at the South African Astronomical Observatory (SAAO) and wish to thank them for their kind help and accommodations. Travel to the SAAO was funded by Australian Government ANSTO AMNRF grant number 09/10-O-03. We thank the referee (D. Uro{\v s}evi\'c) for numerous helpful comments that have greatly improved the quality of this paper.
}

\vskip.6cm


\references

\newcommand{\MNRAS}{\journal{Mon. Not. R. Astron. Soc.}}
\newcommand{\ApJ}{\journal{Astrophys. J.}}
\newcommand{\ApJS}{\journal{Astrophys. J. Supplement}}
\newcommand{\AJ}{\journal{Astronomical. J.}}

\vskip-1mm

Blair, W.~P., Ghavamian, P., Sankrit, R., Danforth, C.~W.: 2006, \ApJS, \vol{165}, 480. 

Boji{\v c}i{\'c}, I.~S., Filipovi{\'c}, M.~D., Parker, Q.~A., Payne, J.~L., Jones, P.~A., Reid, W., Kawamura, A., Fukui, Y.: 2007, \MNRAS, \vol{378}, 1237.

\v{C}ajko K.~O., Crawford E.~J., Filipovi{\'c}, M.~D.: 2009, \journal{Serb. Astron. J.}, \vol{179}, 55. 

Chu, Y-H., Kennicutt, R.~C.: 1988, \AJ, \vol{96}, 1874.

Clarke, J.~N., Little, A.~G., Mills, B.~Y.: 1976, \journal{Aust. J. Phys. Astrophys. Suppl.}, \vol{40}, 1. 

Crawford, E.~J., Filipovi{\'c}, M.~D. and Payne, J.~L.: 2008a, \journal{Serb. Astron. J.}, \vol{176}, 59. 

Crawford, E.~J., Filipovi{\'c}, M.~D., De Horta, A.~Y., Stootman, F.~H., Payne J.~L.: 2008b, \journal{Serb. Astron. J.}, \vol{177}, 61.

Crawford, E.~J., Filipovi{\'c}, M.~D., Haberl, F., Pietsch, W., Payne, J.~L., De Horta, A.~Y.: 2010, \journal{Astron. Astrophys.}, \vol{518}, A35. 

Davies, R.D., Elliott, K. H., Meaburn, J.: 1976, \journal{Mon. Mem. Royal Astron. Society}, \vol{81}, 89.

Di Benedetto, G. P.: 2008, \MNRAS, \vol{390}, 1762.

Filipovi\'c, M.~D., Haynes, R.~F., White, G.~L., Jones, P.~A., Klein, U., Wielebinski, R.: 1995, \journal{Astron. Astrophys. Suppl. Series}, \vol{111}, 331.

Filipovi\'c, M.~D., White, G.~L., Haynes, R.~F., Jones, P.~A., Meinert, D., Wielebinski, R., Klein, U.: 1996, \journal{Astron. Astrophys. Suppl. Series}, \vol{120}, 77. 

Filipovi\'c, M.~D., Haynes, R.~F., White, G.~L., Jones, P.~A.: 1998a, \journal{Astron. Astrophys. Suppl. Series}, \vol{130}, 421.

Filipovi\'c, M.~D., Pietsch, W., Haynes, R.~F., White, G.~L., Jones, P.~A., Wielebinski, R., Klein, U., Dennerl, K., Kahabka, P., Lazendi{\'c}, J.~S.: 1998b, \journal{Astron. Astrophys. Suppl. Series}, \vol{127}, 119.

Filipovi\'c M.~D., Crawford E.~J., Hughes A., Leverenz H., de Horta A.~Y., Payne J.~L., Staveley-Smith L., Dickel J.~R., Stootman F.~H., White G.~L.: 2009, in van Loon J.~T., Oliveira J.~M., eds, \journal{IAU Symposium Vol. 256 of IAU Symposium}, p. PDF8.

Gooch, R.: 2006, Karma Users Manual, ATNF, Sydney.

Haberl, F., Pietsch, W.: 1999, \journal{Astron. Astrophys. Suppl. Series}. \vol{139}, 277.

Heavens, A. F., Meisenheimer, K.: 1987, \MNRAS, \vol{225}, 335.

Hughes, A., Staveley-Smith, L., Kim, S., Wolleben, M., Filipovi{\'c}, M.~D.: 2007, \MNRAS, \vol{382}, 543.

Long, K.~S., Helfand, D.~J., Grabelski, D.~A.: 1981, \ApJ, \vol{248}, 925.

Longair, M. S.: 2000, High Energy Astrophysics (vol. 2), Cambridge University Press.

Mathewson, D.~S., Ford, V.~L., Dopita, M.~A., Tuohy, I.~R., Long, K.~S., Helfand, D.~J.: 1983, \ApJS, \vol{51}, 345.

Mills, B.~Y., Turtle, A.~J., Little, A.~G., Durdin, J.~M.: 1984, \journal{Aust. J. Phys.}, \vol{37}, 321.

Payne, J.~L., White, G.~L., Filipovi{\'c}, M.~D.: 2008, \MNRAS, \vol{383}, 1175.

Rolfs, K., Wilson, T.: 2003, "Tools of Radio Astronomy 4ed.", Springer, Berlin. 

Sault, R.~J., Killeen, N.: 2010, Miriad Users Guide, ATNF, Sydney.

Sault, R.~J., Wieringa, M.~H.: 1994, \journal{Astron. Astrophys. Suppl. Series}, \vol{108}, 585.

Smith, C., Points, S., Winkler, P. F.: 2006, \journal{NOAO Newsletter}, \vol{85}, 6.

\endreferences

}

\end{multicols}

\vfill\eject

{\ }



\naslov{MULTIFREKVENCIONA POSMATRA{NJ}A OSTATAKA SUPERNOVIH U VELIKOM MAGELANOVOM OBLAKU --
SLUQAJ {\bf \SNR} (DEM L204)} 


\authors{L.M. Bozzetto$^1$, M.D.~Filipovi\'c$^1$, E.J. Crawford$^1$, I.S. Boji{\v c}i{\'c}$^1$, J.L. Payne$^1$}\authors{A. Mendik$^1$, B. Wardlaw$^1$ and A.Y. De Horta$^1$}

\vskip3mm


\address{$^1$School of Computing and Mathematics, University of Western 
Sydney\break Locked Bag 1797, Penrith South DC, NSW 1797, Australia} 

\Email{m.filipovic@uws.edu.au}

\vskip3mm


\centerline{\rrm UDK \udc}

\vskip1mm

\centerline{\rit Originalni nauqni rad}

\vskip.7cm

\begin{multicols}{2}

{


\rrm 
U ovoj studiji predstav{lj}amo nove {\rm ATCA} rezultate posmatra{nj}a u radio-kontinumu za ostatak supernove u Velikom Magelanovom Oblaku -- \textrm{\SNR}. Ovaj objekat je tipiqan ostatak supernove sa potkoviqastom morfologijom. Izmerena vrednost dijametra iznosi \mbox{{\rm D=(66$\times$58)$\pm$1}} parseka. Ovo je jedan od najve{\cc}ih ostataka supernovih u Velikom Magelanovom oblaku. Dimenzije ostatka ukazuju da je to stariji objekt, dok je spektralni indeks (\mbox{$\alpha=-0.92\pm$0.11}) veoma strm i povremeno se sre{\cc}e kod mla{dj}ih ostataka. Detektovali smo visok stepen polarizacije qak i do 54\%$\pm$17\% (merenja na 6-cm)."

}\end{multicols}

\end{document}